\documentclass[prl,aps,twocolumn,showpacs]{revtex4}

\usepackage{epsfig}

\begin{document}

\title{Coexistence of Antiferromagnetism and Superconductivity in Electron-doped High-Tc Superconductors} 

\author{W. Yu}
  \email{weiqiang@umd.edu}
\author{J. S. Higgins}
\author{P. Bach}
\author{R. L. Greene}
\affiliation{Center for Superconductivity Research, Department of Physics, \\
University of Maryland, College Park, MD 20742}

\date{\today}
\pacs{74.25.Fy, 73.43.Qt, 74.72.-h, 74.78.Bz}

\begin{abstract}

We present magnetotransport evidence for antiferromagnetism in films of the electron-doped cuprates Pr$_{2-x}$Ce$_x$CuO$_4$. Our results show clear 
signature of static antiferromagnetism up to optimal doping x=0.15, with a quantum phase transition close to x=0.16, and a coexistence of static 
antiferromagnetism and superconductivity for 0.12$\le$x$\le$0.15.      
\end{abstract}

\maketitle

In strongly correlated electron systems, quantum fluctuations close to a quantum critical point lead to many exotic properties of matter 
\cite{Sachdev_science, Coleman_nature}. One example is the unconventional superconductivity (SC) and the non-Fermi liquid normal state properties, 
which appear close to a quantum phase transition (QPT). Such phenomena are found in many heavy Fermion \cite{Paglione, Mathur_nature} and organic 
\cite{Lefebver_prl_2000} superconductors. However, attempts to apply quantum phase transition ideas to describe the properties of the high-T$_C$ 
cuprate superconductors are controversial. In the hole-doped (p-type) cuprates, whether a superconducting fluctuation scenario 
\cite{Emery_nature_374_434} or a competing order scenario \cite{Tallon, Balakirev_nature, Panagopoulos} is an appropriate description of the pseudogap 
phenomena is still highly debated. In the electron-doped (n-type) cuprates, the existence of an antiferromagnetic to paramagnetic QPT is more 
plausible, but there is significant disagreement over if, and where, it occurs and its role in the physical properties \cite{Dagan_prl_92_167001, 
Dagan_prl_94_057005, Li_PRB_75_020506, Uefuji, Kang_nature_423_522, Mang_prb_70_094507, Kang_prb_71_214512, Fujita, Armitage_prl_88_257001, 
Onose_prb_69_024504, Zimmers_EPL_70_225, Motoyama_Nature}.

Several transport studies \cite{Dagan_prl_92_167001, Dagan_prl_94_057005, Li_PRB_75_020506} on electron-doped Pr$_{2-x}$Ce$_x$CuO$_4$ (PCCO) thin 
films suggest an antiferromagnetic QPT inside the superconducting dome at x$\approx$0.16, which is slightly above the optimal doping. Angle resolved 
photoemission spectroscopy (ARPES) measurements on Nd$_{2-x}$Ce$_x$CuO$_4$ (NCCO) \cite{Armitage_prl_88_257001} and optical measurements on NCCO and 
PCCO \cite{Onose_prb_69_024504, Zimmers_EPL_70_225} revealed a normal-state gap which still exists at the optimal doping x=0.15. However, a recent 
inelastic neutron scattering (INS) measurement on NCCO single crystals suggests that long-range order antiferromagnetism (LROAF) does not coexist with 
SC and an antiferromagnetic QPT occurs just before the superconducting dome at x$\approx$0.13 \cite{Motoyama_Nature}. A recent ARPES work on 
superconducting Sm$_{2-x}$Ce$_x$CuO$_4$ (SCCO) single crystals suggests a short-range order antiferromagnetism (SROAF) instead at x=0.14 
\cite{Park_prb_75_060501}. In principle, neutron scattering (NS) and $\mu$SR could differentiate these different interpretations. But, so far, 
measurements from different groups are in significant disagreement \cite{Luke_prb_42_1990, Kang_nature_423_522, Mang_prb_70_094507, Uefuji, 
Motoyama_Nature, Kang_prb_71_214512}. The major experimental difficulty is likely caused by a high-temperature oxygen annealing, which is necessary to 
achieve superconductivity on the n-type cuprates, but also leads to spurious phases \cite{Mang_prb_70_094507} or doping inhomogeneity/uncertainty 
\cite{Kang_prb_71_214512} in large crystals. The controversy over the magnetic properties at high dopings, i.e., x$\ge$0.13, leads to question the 
nature of the QPT proposed by the transport and optical measurements.

\begin{figure}
\includegraphics[width=8cm, height=7cm]{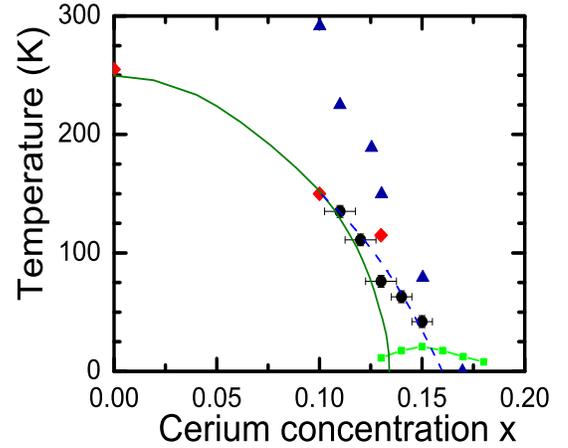}
\caption{\label{PD} The doping dependence of the onset temperature of static antiferromagnetism, T$_A$ (solid circles), determined from angular 
magnetoresistance measurements on our Pr$_{2-x}$Ce$_x$CuO$_4$ films. The dotted line is a guide to the eye. The N\'{e}el temperature T$_N$ of 
Nd$_{2-x}$Ce$_x$CuO$_4$ crystals determined by $\mu$SR \cite{Luke_prb_42_1990} (solid diamonds) and by inelastic neutron scattering  
\cite{Motoyama_Nature}(solid line), and the normal-state gap onset temperature (solid triangles) determined by the optical measurements 
\cite{Onose_prb_69_024504, Zimmers_EPL_70_225} are also shown. The solid squares represent the superconducting transition temperature T$_C$ of our 
films. }
\end{figure}

In this report, we present an in-plane angular magnetoresistance (AMR) study of our PCCO thin films. A fourfold oscillation of the AMR, which is 
caused by the non-collinear antiferromagnetic structure in the n-type cuprate \cite{Lavrov_prl_92_227003}, is used as an indirect method to track the 
AFM ordering. The onset temperature of the fourfold AMR, T$_A$, as shown in Fig.~\ref{PD}, coincides with the N\'{e}el temperature $T_N$  of 
nonsuperconducting NCCO single crystals \cite{Motoyama_Nature, Luke_prb_42_1990} at low dopings (x$\le$0.12) as expected, but deviates from the recent 
INS measurements \cite{Motoyama_Nature} of $T_N$ at high dopings (x$\ge$0.12). Interestingly, T$_A$ extrapolates to zero as x$\rightarrow$0.16, which 
is consistent with the Hall \cite{Dagan_prl_92_167001} and the thermopower \cite{Li_PRB_75_020506} signature of a quantum phase transition at the same 
doping. We believe the fourfold AMR is associated with static antiferromagnetism, and therefore a magnetic origin of the quantum phase transition at 
x$\approx$0.16 is suggested by our new results. Our phase diagram also indicates a coexistence of static antiferromagnetism and superconductivity in 
the doping range 0.12$\le$x$\le$0.15, which suggests that the superconductivity originates from a magnetic mechanism.

\begin{figure}
\includegraphics[width=9cm, height=6cm]{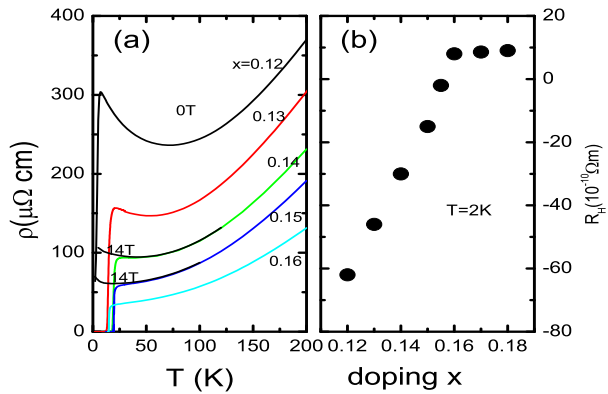}
\caption{\label{RTX} (a) The resistivity of x=0.12, 0.13, 0.14, 0.15, and 0.16 films at zero field and at $\mu _0$H = 14T$\parallel$c-axis. (b) The 
Hall coefficient of 0.12$\le$x$\le$0.18 films (T=2K).}
\end{figure}

Our c-axis oriented PCCO films were prepared by pulsed laser deposition on SrTiO$_3$ or LaSrGaO$_4$  substrates and were reduced {\it in situ} under 
optimized conditions \cite{Maiser_PC}. The superconducting transition temperatures T$_C$, determined by ac susceptibility, are shown in Fig.~\ref{PD}. 
All samples are of typical thickness 3000$\AA$, patterned into a Hall-bar shape, and measured in a Quantum Design Physical Property Measurement System 
(PPMS). The resistivity and Hall coefficient at a few dopings are shown in Fig.~\ref{RTX}. Two features are clearly seen, both of which have been the 
subjects of many studies in the literature. First, a low-temperature resistivity upturn occurs in the normal state at low dopings and disappears at 
x$>$0.16. The antiferromagnetic transition temperature is higher than the upturn temperature at low dopings, but it has been suggested that the upturn 
is related to the AFM \cite{Dagan_prl_94_057005}. Second, the low-temperature Hall data show two different doping dependences, which are connected by 
a kinklike feature at x=0.16 \cite{Dagan_prl_92_167001}. The evolution of the Hall data with doping is consistent with a proposed spin-density-wave to 
paramagnetic QPT at x$\approx$0.16 \cite{Millis}.

\begin{figure}
\includegraphics[width=7cm, height=6cm]{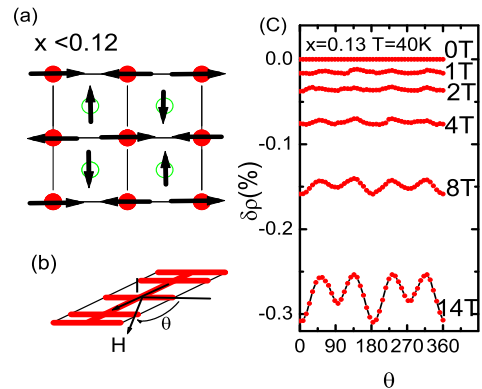}
\caption{\label{RAHx13}(a) The noncollinear antiferromagnetic structure of the n-type cuprates. Solid and hollow circles represent the Cu sites in two 
adjacent planes respectively, and arrows indicate the orientation of the magnetic moments (parallel to the lattice a-axis). (b) The Hall bar pattern 
of films and the ab-plane magnetic field for the angular magnetoresistance (AMR) measurements. $\theta$ is defined as the angle between the magnetic 
field and the Hall bridge (parallel to the lattice a-axis). (c) The AMR, $\delta \rho (\theta, H) =(\rho (\theta, H)-\rho (\theta, H=0))/\rho (\theta, 
H=0)$, for an x=0.13 film at different fields at T=40K. }
\end{figure}

Many underdoped n-type cuprates have a noncollinear antiferromagnetic structure \cite{Skanthanumar_prb_47_6173} below the N\'{e}el temperature, as 
represented in Fig.~\ref{RAHx13}a. A fourfold AMR of the in-plane transport has been found in highly underdoped, antiferromagnetic 
Pr$_{1.29}$La$_{0.7}$Ce$_{0.01}$CuO$_4$ crystals \cite{Lavrov_prl_92_227003}. This is caused by a strong spin-orbit coupling and an anisotropic 
(fourfold) spin-flop field, with the easy-axis along the lattice diagonal direction and the hard-axis along the lattice a-axis 
\cite{Lavrov_prl_92_227003}. In this paper, we focus on our in-plane angular magnetoresistance studies on PCCO films, in particular at high dopings. 
Our films are mounted on a rotator and the sample rotates around the lattice c-axis with the magnetic field confined in the ab-plane (see 
Fig.~\ref{RAHx13}b). The AMR of an x=0.13 superconducting film, $\delta \rho (\theta, H)$, is shown at fields up to 14T at 40K in Fig.~\ref{RAHx13}c. 
With increasing field, a small modulation of the AMR develops. At $\mu _0$H = 14T, a fourfold oscillation is clearly seen.

\begin{figure}
\includegraphics[width=9cm, height=12cm]{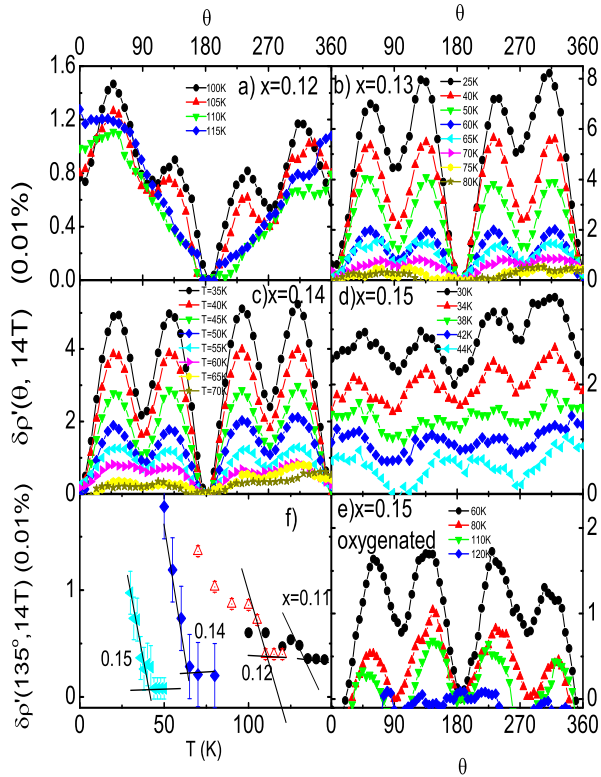}
\caption{\label{RATX} (a)-(e), The temperature dependence of the relative AMR ($\delta \rho' (\theta, 14T)=\delta \rho (\theta, 14T)-\delta \rho 
(\theta=180^\circ, 14T)$) for optimally oxygen-reduced x=0.12, x=0.13, x=0.14, and x=0.15 PCCO films, and an oxygenated x=0.15 PCCO film at 14T. Plots 
in d) are shifted vertically for clarity. (f), The temperature dependence of $\delta \rho' (\theta, 14T)$ of several films with the magnetic field 
along the lattice diagonal direction $\theta = 135^\circ$.}
\end{figure}

A fourfold oscillation is found to exist below a specific temperature in all films with doping up to x=0.15. In Fig.~\ref{RATX}a-d, the relative AMR, 
$\delta \rho' (\theta, 14T)$, is plotted to show the temperature dependence of the AMR modulation for several dopings. For each doping, the fourfold 
oscillation of the AMR emerges below an onset temperature T$_A$. In Fig.~\ref{RATX}f, the temperature dependence of the magnetoresistance with field 
along the lattice diagonal direction $\theta = 135^\circ$ (easy-axis) is plotted. $\delta \rho' (\theta, 14T)$ also shows a kinklike behavior at T$_A$ 
for each doping. The values of T$_A$ for films 0.11$\le$x$\le$0.15 are plotted in Fig.~\ref{PD}. T$_A$ decreases from 135K to 65K as doping increases 
from x=0.11 to 0.14. For optimal doping x=0.15, a fourfold pattern is also clearly seen at T = 30K and not discernible above T=44K. For overdoped 
x=0.155 and x=0.16 films (data not shown), the fourfold oscillation is not seen above T$_C$. At present, we cannot measure AMR in the superconducting 
samples below T$_C$ since the in-plane $H_{C2}$ is too large. Therefore, if the fourfold oscillation exists, its onset temperature is below $T_C$. 

For comparison, the N\'{e}el temperature T$_N$ of non-superconducting NCCO crystals from $\mu$SR \cite{Luke_prb_42_1990} and from the INS 
\cite{Motoyama_Nature} measurements is also plotted in Fig.~\ref{PD}. We also show the temperature T$_P$, below which a partial gap in $\sigma_{ab}$ 
is seen in the optical measurements \cite{Onose_prb_69_024504, Zimmers_EPL_70_225}. The fact that T$_P$ is higher than T$_N$ has been attributed to 
antiferromagnetic fluctuations \cite{Onose_prb_69_024504, Zimmers_EPL_70_225}. For underdoped, non-superconducting (x$\le 0.12$) films, T$_A$ is much 
lower than T$_P$, but consistent with the value of T$_N$. In Fig.~\ref{RATX}e, we also show the AMR data for an oxygenated, non-superconducting 
Pr$_{1.85}$Ce$_{0.15}$CuO$_4$ film, which has a resistivity similar to as-grown Pr$_{1.85}$Ce$_{0.15}$CuO$_4$ crystals. Our low-temperature AMR data 
is consistent with an earlier AMR study on as-grown x=0.15 crystals \cite{Fournier_prb_69_220501}. Both the amplitude and the onset temperature of the 
fourfold AMR are much higher than that of the superconducting x=0.15 film (Fig.~\ref{RATX}d). The fourfold AMR emerges below an onset temperature 
T$_A$$\approx$115K, again consistent with the N\'{e}el temperature determined by elastic neutron scattering of as-grown crystals 
\cite{Mang_prl_93_027002}.  {\it These facts strongly suggest that the fourfold AMR at low dopings is indicative of LROAF, and not antiferromagnetic 
fluctuations}. 

For higher doping $0.13 \le$x$\le$0.15, no LROAF is found in superconducting NCCO crystals by the INS measurement\cite{Motoyama_Nature}, which seems 
to be inconsistent with our finite T$_A$. We believe our fourfold AMR is not due to the spurious magnetic oxide phase  \cite{Mang_prb_70_094507} nor 
doping inhomogeneity induced by oxygen reduction. The large increase under oxygenation of the AMR amplitude and the AMR onset temperature (see 
Fig.~\ref{RATX}d and e), rules out a spurious phase contribution. Compared to bulk NS crystals, our films appear to have a good control of the oxygen 
as indicated by the sharp superconducting transitions found by the ac susceptibility ($\Delta T_C < 1K$ for all films). For our x$\ge$0.1 films, we 
believe the doping inhomogeneity/uncertainty is below $\Delta x=0.005$ due to our very well controlled growth and oxygen annealing conditions. 
Interestingly, our T$_A$ extrapolated to zero at x$\approx$0.16. In comparison with the Hall \cite{Dagan_prl_92_167001}, thermopower 
\cite{Li_PRB_75_020506}, and ARPES \cite{Armitage_prl_88_257001} measurements, our AMR gives further support for a QPT of magnetic origin at this 
doping.

In order to understand our data compared with the INS data, we suggest a few possible explanations below. 

First, the actual carrier density may be different for the INS single crystals and our AMR thin films, due to different oxygen annealing conditions. 
Therefore, the fourfold AMR in PCCO could be caused by long-range ordering and the QPT at x=0.16 is associated with a LROAF. However, this would 
indicate a shift of doping by x$\approx$0.03 between our films and the INS crystals, which seems to be unlikely. The INS suggests no coexistence of 
LROAF and SC, which seems to be inconsistent with our observation of fourfold AFM up to optimal doping. It is possible that PCCO has a long-range AFM 
QPT inside the superconducting dome, while this is not the case for NCCO \cite{xdif}. However, this is unlikely because both systems have a similar 
AFM transition temperature at low dopings and a Fermi surface topological reconstruction at x$\approx$0.16 \cite{Dagan_prl_92_167001, 
Armitage_prl_88_257001}. We believe that INS and Hall (or thermopower) measurements on the same NCCO single crystal should be able to clarify these 
two scenarios. 

Second, The fourfold AMR method may not be able to distinguish between a long-range order antiferromagnetism and a {\it static} short-range order 
antiferromagnetism. Considering the disappearance of the LROAF at x$\ge$0.13 as shown by the INS data, our fourfold AMR may suggest a SROAF at higher 
dopings. Indeed, signatures of SROAF have been shown by INS \cite{Motoyama_Nature}, although this could arise from oxygen inhomogeneity in the larger 
crystals used for neutron scattering. Recent ARPES measurements on SCCO single crystals are also suggestive of such a SROAF scenario 
\cite{Park_prb_75_060501}. Therefore, the disappearance of T$_A$ at x$\approx$0.16 might suggest that the nature of the QPT at x$\approx$0.16, as 
revealed by the Hall, resistivity, and thermopower measurements, is related to a SROAF.

Third, the fourfold AMR could also be caused by a {\it quasi-static} SROAF, i.e., a fluctuating order, if finite disorder pins the slow fluctuation 
with a time scale sufficient for transport measurements. This could be the case for the optimal-doped x=0.15 films, where a much smaller fourfold AMR 
is observed (Fig.~\ref{RATX}d). 

Surprisingly, the dramatic change of the the Fermi surface at x$\approx$0.16, as shown by the ARPES \cite{Armitage_prl_88_257001} and the transport 
\cite{Dagan_prl_92_167001, Li_PRB_75_020506} measurements, suggests that the Fermi surface topological reconstruction is strongly affected by this 
SROAF rather than the LROAF. For comparison, the Fermi surface evolution is usually associated with a long range ordering in most magnetic systems 
\cite{Yeh_nature}. Therefore, our data may suggest a new type of QPT with different properties from the one found in the INS experiments.    

The short-range ordering phase seems to be similar to a high-pressure partial order phase beyond a first order phase transition in MnSi 
\cite{MnSi,Yu_MnSi}. Unfortunately, the amplitude of the AMR is not simply linear in the amplitude of the order parameter, as indicated by the 
temperature dependence of the fourfold AMR (see Fig.~\ref{RATX}f). Therefore, we are not able to distinguish between a first order and a second order 
phase transition by the AMR measurements. Other scenarios, such as a spin glass transition \cite{wilson_prb_74_144514} or stripes 
\cite{Fournier_prb_69_220501} are also possible. Although these phases are supported in some p-type cuprates \cite{Panagopoulos, 
Tranquada_Nature_1995}, strong evidence for either of these has not been found in the n-type cuprates. 

Our AMR evidence of static antiferromagnetism with a QPT inside the superconducting dome also provides a framework to understand other experimental 
results in the n-type cuprates. First, our T$_{A}$ is slightly above the resistivity upturn temperature, which supports the view that the upturn is  
related to the static AFM \cite{Dagan_prl_94_057005}, although more studies are required to understand this in detail. Our measurement also suggests a 
coexistence of SROAF and SC up to optimal doping x=0.15. The dome-like doping dependence of T$_C$ in the underdoped regime may be naturally explained 
by a competition between the coexisting SROAF and SC.   
   
In summary, we determined the doping dependence of static (or quasi-static) antiferromagnetism in the electron-doped cuprates by an angular 
magnetoresistance method on Pr$_{2-x}$Ce$_x$CuO$_4$ films. Our data give evidence for the existence of intrinsic static antiferromagnetism up to 
x=0.15, which is consistent with the proposed quantum phase transition at x$\approx$0.16 \cite{Dagan_prl_92_167001, Dagan_prl_94_057005, 
Li_PRB_75_020506}. Compared with the inelastic neutron scattering evidence \cite{Motoyama_Nature} for a long-range order antiferromagnetic quantum 
phase transition at x$\approx$0.13, our angular magnetoresistance measurements suggest a short-range order antiferromagnetic quantum phase transition 
at x$\approx$0.16.  

This work is supported by NSF under Contract No. DMR-0352735. The authors would acknowledge fruitful discussions with J. W. Lynn, N. P. Armitage, and 
S. E. Brown. 


\begin{thebibliography}{35}
\bibitem[{xdi()}]{xdif}
\bibinfo{note}{Here we limit our discussion to the n-type cuprates that have an
  optimal T$_C$ at cerium doping x=0.15. Some La-based n-type cuprates
  \cite{wilson_prb_74_144514} are known to have a lower optimal cerium doping.}
\end{thebibliography}

\end{document}